\documentclass[superscriptaddress,aps,preprintnumbers,amsmath,amssymb,prd,nofootinbib,preprint]{revtex4-1}
\pdfoutput=1
\usepackage{graphicx}
\usepackage{epstopdf}
\usepackage{dcolumn}% Align table columns on decimal point
\usepackage{bm}% bold math
\usepackage{hyperref}
\usepackage{color}
\usepackage{setspace}

\begin{document}

%%%%%%%%%%%%%%%%%%%%%%%%%%%%%%%%%%%%%%%%%%%

\def\a{\alpha}
\def\b{\beta}
\def\c{\varepsilon}
\def\d{\delta}
\def\e{\epsilon}
\def\f{\phi}
\def\g{\gamma}
\def\h{\theta}
\def\k{\kappa}
\def\l{\lambda}
\def\m{\mu}
\def\n{\nu}
\def\p{\psi}
\def\q{\partial}
\def\r{\rho}
\def\s{\sigma}
\def\t{\tau}
\def\u{\upsilon}
\def\v{\varphi}
\def\w{\omega}
\def\x{\xi}
\def\y{\eta}
\def\z{\zeta}
\def\D{\Delta}
\def\G{\Gamma}
\def\H{\Theta}
\def\L{\Lambda}
\def\F{\Phi}
\def\P{\Psi}
\def\S{\Sigma}

\def\o{\over}
\def\beq{\begin{eqnarray}}
\def\eeq{\end{eqnarray}}
\newcommand{\gsim}{ \mathop{}_{\textstyle \sim}^{\textstyle >} }
\newcommand{\lsim}{ \mathop{}_{\textstyle \sim}^{\textstyle <} }
\newcommand{\vev}[1]{ \left\langle {#1} \right\rangle }
\newcommand{\bra}[1]{ \langle {#1} | }
\newcommand{\ket}[1]{ | {#1} \rangle }
\newcommand{\EV}{ {\rm eV} }
\newcommand{\KEV}{ {\rm keV} }
\newcommand{\MEV}{ {\rm MeV} }
\newcommand{\GEV}{ {\rm GeV} }
\newcommand{\TEV}{ {\rm TeV} }
\newcommand{\1}{\mbox{1}\hspace{-0.25em}\mbox{l}}
\newcommand{\headline}[1]{\noindent{\bf #1}}
\def\diag{\mathop{\rm diag}\nolimits}
\def\Spin{\mathop{\rm Spin}}
\def\SO{\mathop{\rm SO}}
\def\O{\mathop{\rm O}}
\def\SU{\mathop{\rm SU}}
\def\U{\mathop{\rm U}}
\def\Sp{\mathop{\rm Sp}}
\def\SL{\mathop{\rm SL}}
\def\tr{\mathop{\rm tr}}
\def\mpl{M_{PL}}

\def\IJMP{Int.~J.~Mod.~Phys. }
\def\MPL{Mod.~Phys.~Lett. }
\def\NP{Nucl.~Phys. }
\def\PL{Phys.~Lett. }
\def\PR{Phys.~Rev. }
\def\PRL{Phys.~Rev.~Lett. }
\def\PTP{Prog.~Theor.~Phys. }
\def\ZP{Z.~Phys. }

\def\dd{\mathrm{d}}
\def\ff{\mathrm{f}}
\def\BH{{\rm BH}}
\def\inf{{\rm inf}}
\def\ev{{\rm evap}}
\def\eq{{\rm eq}}
\def\SM{{\rm sm}}
\def\Mpl{M_{\rm Pl}}
\def\GeV{{\rm GeV}}
\newcommand{\Red}[1]{\textcolor{red}{#1}}

% To increase line space in footnote.
%

%%%%%%%%%%%%%%%%%%%%%%%%%%%%%%%%%%%%%%%%%%%%%%%%%%%%%%%%%%%%%%%

\title{Diboson Resonance as a Portal to Hidden Strong Dynamics}
\author{Cheng-Wei Chiang}
\affiliation{Center for Mathematics and Theoretical Physics and Department of Physics, National Central University, Taoyuan, Taiwan 32001, R.O.C.}
\affiliation{Institute of Physics, Academia Sinica, Taipei, Taiwan 11529, R.O.C.}
\affiliation{Physics Division, National Center for Theoretical Sciences, Hsinchu, Taiwan 30013, R.O.C.}
\affiliation{Kavli IPMU (WPI), UTIAS, University of Tokyo, Kashiwa, Chiba 277-8583, Japan}
\author{Hajime Fukuda}
\affiliation{Kavli IPMU (WPI), UTIAS, University of Tokyo, Kashiwa, Chiba 277-8583, Japan}
\author{Keisuke Harigaya}
\affiliation{ICRR, University of Tokyo, Kashiwa, Chiba 277-8582, Japan}
\author{Masahiro Ibe}
\affiliation{Kavli IPMU (WPI), UTIAS, University of Tokyo, Kashiwa, Chiba 277-8583, Japan}
\affiliation{ICRR, University of Tokyo, Kashiwa, Chiba 277-8582, Japan}
\author{Tsutomu T. Yanagida}
\affiliation{Kavli IPMU (WPI), UTIAS, University of Tokyo, Kashiwa, Chiba 277-8583, Japan}

\begin{abstract}
We propose a new explanation for excess events
observed in the search
for a high-mass resonance decaying into dibosons by the ATLAS experiment.
The resonance is identified as a {\it composite} spin-$0$ particle that couples
to the Standard Model gauge bosons via dimension-5 operators.
The excess events can be explained if the dimension-5 operators
are suppressed by a mass scale of ${\cal O}(1$--$10$)\,TeV.
We also construct a model of hidden strong gauge dynamics which realizes the 
spin-$0$ particle as its lightest composite state,
with appropriate couplings to Standard Model gauge bosons.
\end{abstract}

\date{\today}
\maketitle
\preprint{IPMU15-0104}
%%%%%%%%%%%%%%%%%%%%%%%%%%%%%%%%%%%%%%%%%%%%%%%%%%%%%%%%%%%%%%%

\section{Introduction}
Recently, the ATLAS Collaboration reported excess events in the search for a high-mass resonance decaying
into dibosons which subsequently decay hadronically~\cite{Aad:2015owa}. 
The excess peaks at the diboson invariant mass around $2$\,TeV.
For an integrated luminosity of $20\,$fb$^{-1}$ in the 8-TeV LHC run,
the local significances of the excess events are 
$3.4\,\s$,
$2.6\,\s$,
and $2.9\,\s$
when they are interpreted as the decay of the resonance into $WZ$, $WW$ and $ZZ$, respectively.

Inspired by the report, quite a few explanations have been proposed so far.
Most of such studies employ new spin-$1$ resonances, 
such as $W'/Z'$  bosons in extended electroweak 
gauge sectors~\cite{Hisano:2015gna,Cheung:2015nha,Dobrescu:2015qna,Alves:2015mua,Gao:2015irw,Thamm:2015csa,Brehmer:2015cia,Cao:2015lia,Cacciapaglia:2015eea,Abe:2015jra,Abe,Allanach},%
\footnote{
See Ref.~\cite{Abe} for prospects of the searches for $W'/Z'$ resonances at the LHC Run-II.}
a massive spin-$1$ boson in composite Higgs models~\cite{Fukano:2015hga,Carmona}, 
or a phenomenological vector particle model~\cite{Franzosi:2015zra}. 
A possible explanation by a heavy Higgs boson is also mentioned in Ref.~\cite{Aguilar-Saavedra:2015rna}.  

In this paper, we propose another promising possibility where the resonance is identified 
as a {\it composite} spin-0 neutral particle that couples to the Standard Model (SM) gauge bosons via dimension-5 operators.  As we will show, the excess events can be explained if dimension-5 operators are suppressed by a mass scale of ${\cal O}(1$--$10$)\,TeV.

We also construct a model of hidden strong dynamics which realizes the above-mentioned spin-$0$ particle as its lightest composite state, with appropriate couplings to the SM gauge bosons.
In this case, the composite spin-$0$ particle consists of bi-fundamental scalars under the hidden and the SM gauge symmetries.
The mass of the resonance as well as the suppression scale of ${\cal O}(1)$\,TeV
are achieved when the hidden strong dynamics exhibits confinement at a dynamical scale of ${\cal O}(1)$\,TeV.

The organization of the paper is as follows.
In section\,\ref{sec:SEFT}, 
we discuss whether a spin-$0$ resonance can explain the excess events 
when it couples to the SM gauge bosons via dimension-5 operators.
In section\,\ref{sec:model},
we propose a model of hidden strong dynamics which yields a composite spin-$0$ particle
with appropriate couplings to the SM gauge bosons.
Discussions and conclusions are given in the last section.

%%%%%%%%%%%%%%%%%%%%%%%%%%%%%%%%%%%%%%%%%%%%%%%%%%%%%%%%%%%%%%%
\section{Effective Field Theory of Diboson Resonance}
\label{sec:SEFT}

In our proposal, the diboson resonance will eventually be identified as the lightest composite scalar boson 
in a hidden sector with strong dynamics.
The hidden sector couples to the SM sector
through fields charged under both the hidden and the SM gauge symmetries.
We assume that the hidden strong dynamics exhibits confinement
at a dynamical scale around a few TeV, $\Lambda_{\rm dyn} \sim {\cal O}(1)$\,TeV,  
leaving a spin-$0$ particle as the lightest state, which couples 
to the SM gauge bosons via higher dimensional operators.
Before elucidating explicit models of the hidden strong dynamics, let us discuss 
how the observed excess events can be explained by a composite spin-$0$ particle 
using the effective field theory approach.

Let us consider an effective field theory which consists of its lightest neutral scalar boson $S$ and SM particles,
after integrating out heavier degrees of freedom.
% in the hidden strong dynamics. 
Effective interactions of $S$ with the SM gauge bosons are given by 
\begin{eqnarray}
\label{eq:SEFT}
{\cal L}_{\rm eff} =
 \frac{\k_3 }{\L}S G^a_{\mu\nu}G^{a\,\mu\nu}
 + 
  \frac{\k_2}{\L}S W^i_{\mu\nu}W^{i\,\mu\nu}
 + 
\frac{5}{3} \frac{\k_1 }{\L}S B_{\mu\nu}B^{\mu\nu}\ ,
\end{eqnarray}
where $\Lambda$ is a suppression scale.
Here, $G$, $W$ and $B$ denote the field strengths of the SM gauge bosons
of the
$SU(3)_c$, 
$SU(2)_L$, and 
$U(1)_Y$ groups, respectively, with the superscripts $a$ and $i$ being the indices for the corresponding adjoint representations.
%run $a =1-8$ and  $i=1-3$, respectively.
The field strengths are normalized so that their kinetic terms are given by,
\begin{eqnarray}
{\cal L}=- \frac{1}{4g_s^2}  G^a_{\mu\nu}G^{a\,\mu\nu} 
- \frac{1}{4g^2} W^i_{\mu\nu}W^{i\,\mu\nu} 
-\frac{1}{4g^{'2}} B_{\mu\nu}B^{\mu\nu}\ ,
\end{eqnarray}
where $g_s$, $g$ and $g'$ are the corresponding gauge coupling constants. 
The coefficients $\k_{3,2,1}$ are of ${\cal O}(1)$ and encapsulate 
details of the strong dynamics.

\subsection{Production cross section of the scalar resonance}

Through the above effective interactions,  the parton-level production cross section 
of $S$ via the gluon fusion process is given by
\begin{eqnarray}
\hat\sigma(g+g\to S)\simeq \frac{\pi^2}{8 M_S}~ \Gamma(S\to g+g) ~ \delta(\hat s-M_S^2)\ ,
%\hat\sigma(g+g\to S)\simeq \frac{1}{128} \frac{16\pi^2}{M_S^2} \times M_S\Gamma(S\to g+g) \times \delta(\hat s-M_S^2)\ ,
\label{eq:sigmahat}
\end{eqnarray}
in the narrow width approximation, as suggested by the result of ATLAS experiment.
In Eq.~(\ref{eq:sigmahat}), $M_S$ denotes the mass of the scalar boson $S$ and $\hat s$ the square of 
the partonic center-of-mass energy.
The partial decay width of $S$ into a pair of gluons is
\begin{eqnarray}
\label{eq:SGG}
\G(S\to g+g) = \frac{2}{\pi} \left( \frac{g_s^2\k_3}{\Lambda}  \right)^2 M_S^3\ .
\end{eqnarray}

After convolution with the parton distribution function (PDF) of the gluon inside the proton, $f_g$, 
the total production cross section in the proton-proton collision becomes
\begin{eqnarray}
\sigma(p+p\to S) &=&
\frac{\pi^2}{8} \left(\frac{\Gamma(S\to g+g)}{M_S}\right) \times 
\left[\frac{1}{s}\frac{\q{\cal L}_{gg}}{\q\cal \t}
\right]\ ,\cr
\frac{\q{\cal L}_{gg}}{\q\cal \t} &=& \int_0dx_1 dx_2 f_g(x_1) f_g(x_2) \delta (x_1 x_2 - \tau)\ ,
\end{eqnarray}
where $\tau = M_S^2/s$ and $\sqrt{s} = 8$ TeV.
For $M_S \simeq 2$\,TeV, the luminosity function
\begin{eqnarray}
\frac{\q{\cal L}_{gg}}{\q\cal \t} &\simeq&  0.18\  \,\, (0.14) \ ,
\end{eqnarray}
where we have used the PDF's of MSTW2008~\cite{Martin:2009iq} and fixed the factorization scale and the renormalization scale to be $\mu = M_S/2 = 1$\,TeV 
($\mu = M_S = 2$\,TeV).
With $\sqrt{s} = 8$\,TeV, this amounts to
\begin{eqnarray}
\frac{1}{s}\frac{\q{\cal L}_{gg}}{\q\cal \t} &\simeq&  1.1\,{\rm pb}\  \,\, (0.85\,{\rm pb}) \ .
\end{eqnarray}
For comparison, we also give the corresponding values for $\sqrt{s} = 13$\,TeV:
\begin{eqnarray}
\frac{\q{\cal L}_{gg}}{\q\cal \t} \simeq  6.6\  \,\, (5.7) \ , \quad
\frac{1}{s}\frac{\q{\cal L}_{gg}}{\q\cal \t} \simeq  15\,{\rm pb}\  \,\, (13\,{\rm pb}) \ .
\label{eq:13TeV}
\end{eqnarray}
The production cross section of $S$ becomes about ten times larger at LHC Run-II.

%%%%%%%%%%%%%%%%%%%%%%%%
\subsection{The diboson excess at the LHC}
The partial decay widths of the scalar boson $S$ into the other gauge bosons, $W$, $Z$ and $A$ (photon), are given by
\begin{eqnarray}
\label{eq:SWW}
\G(S\to W^+ + W^-) &=& \frac{1}{2} 
 \frac{1}{\pi} \left( \frac{g^2\k_2}{\Lambda} \right)^2 M_S^3\ , \\
 \label{eq:SZZ}
 \G(S\to Z+Z) &=& \frac{1}{4} 
 \frac{1}{\pi} \left[
 \left( \frac{g^2\k_2}{\Lambda}\right)  c_W^2
+
\frac{3}{5}\left(
\frac{g^{\prime 2}\k_1}{\Lambda} 
\right)s_W^2 
 \right]^2 M_S^3\ , 
 \\
 \G(S\to \g+\g) &=& \frac{1}{4} 
 \frac{1}{\pi} \left[ 
  \left( \frac{g^2\k_2}{\Lambda}\right)  s_W^2
+
\frac{3}{5}\left(
\frac{g^{\prime 2}\k_1}{\Lambda} 
\right)
c_W^2
 \right]^2 M_S^3\ , \\
 \G(S\to Z+\g) &=& \frac{1}{2} 
 \frac{1}{\pi} \left[ 
  \left( \frac{g^2\k_2}{\Lambda}\right)  
-
\frac{3}{5}\left(
\frac{g^{\prime 2}\k_1}{\Lambda} 
\right)
 \right]^2
 c_W^2s_W^2
 M_S^3\ , 
\end{eqnarray}
where $s_W \equiv \sin\theta_W$ with $\theta_W$ being the weak mixing angle, $c_W = (1-s_W^2)^{1/2}$, and
the masses of the $W$ and $Z$ bosons are neglected.
It should be noted that the resonance does not decay into SM fermions or Higgs bosons in this model.

As long as $\k_1$ is not much larger than $\k_2$,%
\footnote{This is the case for a model discussed in the next section.} 
the $WW$ and $ZZ$ modes have the dominant partial widths among the four decays.
As a result, the branching ratios of the $WW$, $ZZ$, and $gg$ modes roughly satisfy the following relation:
\begin{eqnarray}
\label{eq:B}
B_{WW} + B_{ZZ}  \simeq 1 - B_{gg}\ .
\end{eqnarray}
%where $B_{WW,ZZ,gg}$ denote the branching ratios into $WW$, $ZZ$ and $gg$, respectively.
Therefore, the total production cross section of the resonance decaying into $WW$ and $ZZ$ channels is
\begin{eqnarray}
\sigma_{WW}  + \sigma_{ZZ} &\simeq&
\frac{16\pi^2}{128} \left(\frac{\Gamma_S}{M_S}\right) B_{gg}(1-B_{gg}) 
\times 
\left[\frac{1}{s}\frac{\q{\cal L}_{gg}}{\q\cal \t}
\right]\ ,
\end{eqnarray}
where $\G_S$ denotes the total decay width of $S$.

In the analysis of Ref.~\cite{Aad:2015owa}, excess is observed in each of the $WZ$, $WW$ and $ZZ$ channels.
%in the search for a high-mass resonance decaying into dibosons
%are interpreted
%as the channels
%as pairs of $WZ$, $WW$ or $ZZ$.
At this point, however, the observed signals can be explained by purely an excess in
$WW$ and/or $ZZ$ with a cross section of ${\cal O}(1$--$10)$\,fb, 
as the experimental selection criteria for the $W$ and $Z$ bosons are not very discriminative.%
\footnote{In fact, roughly $20$\% of the excess events 
can be interpreted either of $WZ$, $WW$ and $ZZ$\,\cite{Aad:2015owa}.
See also Ref.~\cite{Allanach}, which argues that the excess events can be explained with 
$5\,{\rm fb}\lesssim\s_{WW} + \s_{ZZ}\lesssim 20\,$fb.
}
Therefore, it is possible to identify the diboson resonance with the neutral spin-$0$ boson $S$ 
that does not necessarily decay into the $WZ$ final state.

It should be noted that  stringent bounds $\s_{WW}\lesssim 3$--$5$\,fb have been placed by CMS and ATLAS Collaborations
in Refs.\,\cite{Khachatryan:2014gha, Aad:2015ufa} from the semi-leptonic channel searches for $M_S \simeq 2$\,TeV.
One caveat here is, however,  that the $W$ and $Z$ bosons from the decays of $S$ are in the transverse modes.  
This feature can make some slight differences in selection efficiencies from the ones estimated in Ref.~\cite{Aad:2015owa}, 
where the $W$ and $Z$ bosons are assumed to be in the longitudinal modes.
Besides, higher-order QCD corrections to the production cross section
(the so-called $K$-factor), can be sizeable.\footnotemark[4]
With these reasons, we are satisfied with concentrating on the parameter space where the leading order cross section $\sigma_{WW} + \sigma_{ZZ} = {\cal O}(1$--$10)$\,fb.
%as a region which
%is appropriate to
%can explain the excess by the spin-$0$ resonances. 
For a more accurate estimate of the viable parameter range in the effective field theory, we will need higher-order corrections as well as detailed calculations that are beyond the scope of this paper.

%%%%%%%%%%%%%%%%%%%%%%%%%%%%%%%%%%%%%%
\begin{figure}[t]
\begin{center}
  \includegraphics[width=.6\linewidth]{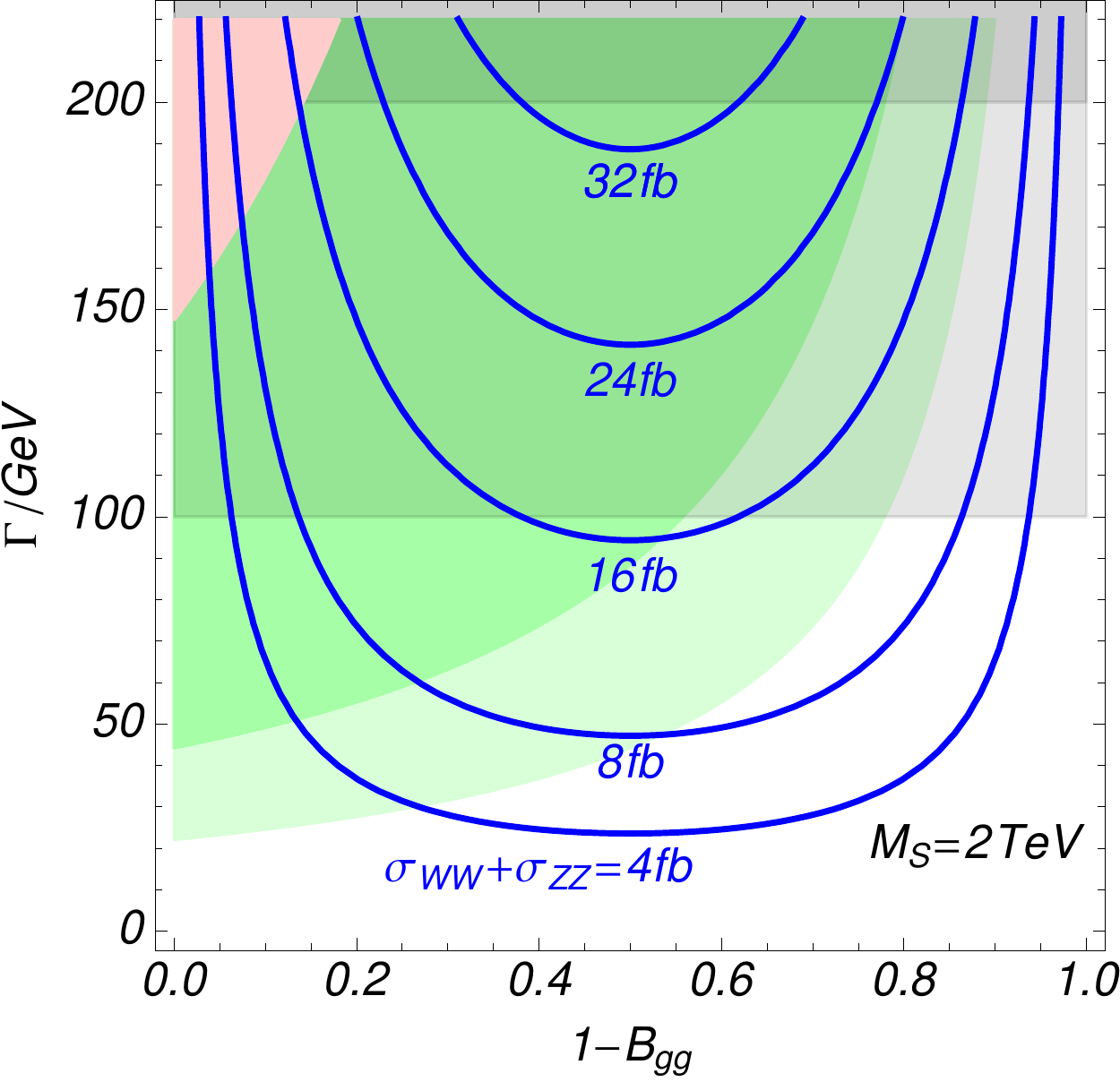}
  \end{center}
\caption{\sl \small
Contours of the production cross section of $S$ times its decay branching ratios into $WW$ and $ZZ$ on the $\G_S$-$(1-B_{\rm gg})$ plane at the 8-TeV LHC.  We fix $M_S = 2$\,TeV and take the factorization and renormalization scale $\mu = M_S/2$.
The lighter (darker) gray region is disfavored by the narrow width assumption $\G_S \lesssim 100$\,GeV ($\alt 200$\,GeV).
The pink shaded region is excluded by the constraint from the dijet channel\,\cite{Aad:2014aqa,Khachatryan:2015sja}.
The  green and light green shaded regions show the constraints from the diphoton channel search,  $\s_{\g\g}\lesssim 0.3$\,fb\,\cite{Aad:2015mna},
for typical branching fractions $B_{\g\g} = 1$\% and $B_{\g\g}=2$\%, respectively.
}
\label{fig:crosssection}
\end{figure}
%%%%%%%%%%%%%%%%%%%%%%%%%%%%%%%%%%%%%%

In Fig.\,{\ref{fig:crosssection}}, we show a contour plot of the total production cross section of $S$ that decays into $WW$ and $ZZ$ pairs on the plane of $1-B_{gg}$ and $\Gamma_S$ for $M_S = 2$\,TeV.
In the figure, the gray region is disfavored by the requirement of a narrow resonance:
$\G_S \lesssim 100$\,GeV for the light gray are and $\alt 200$\,GeV for the darker gray area.
The pink region is excluded by the constraint from the dijet channel, {\it i.e.}, $\sigma(p+p\to S \to g + g)\lesssim 100$\,fb \,\cite{Aad:2014aqa,Khachatryan:2015sja}.
The figure shows that the required cross section of ${\cal O}(1$--$10)$\,fb can be realized 
in the parameter space where $B_{WW}+B_{ZZ} = {\cal O}(10)$\%
without any conflict with the narrow width approximation or the dijet constraint. 

\footnotetext[4]{See {\it e.g.}, Ref.~\cite{Baglio:2010ae} for a discussion on the $K$-factor for the Higgs production,
although their analysis cannot be directly applied to the $S$ production here since the effective field theory is valid up to ${\cal O}(\Lambda)$, whereas the Higgs effective field theory is valid only for a Higgs mass below twice of the top-quark mass.
}

It should be emphasized that the suppressed couplings to the SM fermions are one of the striking features of the
composite scalar resonance where the composite scalar couples to the SM fermions via the mixing to the Higgs bosons.
These features should be compared with $W'/Z'$ bosons or generic composite scalar resonance (see e.g.\,\cite{Barbieri:2010nc}).
Therefore, our model is free from the constraints of dilepton mode searches, 
$\s(p+p\to S \to \ell + \ell')\lesssim 1$\,fb~\cite{ATLAS:2014wra,Khachatryan:2014tva} and the decay into a Higgs 
boson and a $Z$ boson, $\s(p+p\to S \to Z + h) \lesssim 7\,$fb~\cite{Khachatryan:2015bma}. 

%%%%%%%%%%%%%%%%%%%%%%%%%%%%%%%%%%%%%%
\begin{figure}[t]
\begin{center}
  \includegraphics[width=0.8\linewidth]{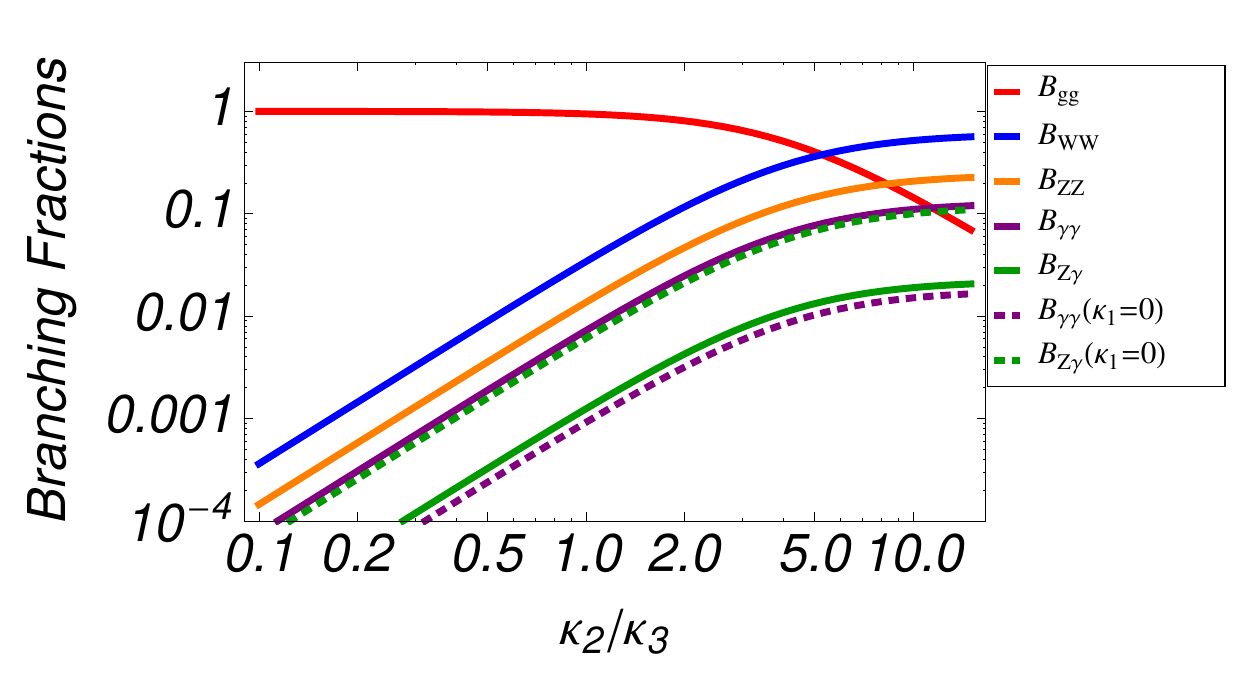}
 \end{center}
\caption{\sl \small
Decay branching ratios of $S$ into different modes as a function of $\k_2/\k_3$.
Solid curves are drawn by assuming $\k_1 = \k_2$.
The branching ratios of the $\g\g$ and $Z\g$ modes for $\k_1 = 0$ are also drawn for comparison,
while the other modes are barely changed when $\k_1 = 0$.}
\label{fig:branching}
\end{figure}
%%%%%%%%%%%%%%%%%%%%%%%%%%%%%%%%%%%%%%

It is also noted that a spin-$0$ resonance can decay into a pair of photons.
This feature should be contrasted with models where the diboson resonance is interpreted as a massive spin-$1$ particle whose decay into a pair of photons is forbidden by the Landau-Yang theorem\,\cite{Landau:1948kw,Yang:1950rg}. 
Therefore, an observation of excess in the diphoton mode will be a smoking gun signal of models with 
a spin-$0$ resonance.
In Fig.\,\ref{fig:branching}, we show the branching ratios of all the allowed diboson modes as a function of $\k_2/\k_3$, the ratio of effective coupling strengths in the weak and strong interactions.
The solid curves are drawn under the assumption that $\k_1 = \k_2$.  We also show the branching ratios of the $\g\g$ and $Z\g$ mode for $\k_1 = 0$ as a comparison.  Therefore, the branching ratio of $S$ decaying into two photons is expected to be ${\cal O}(1$--$10)$\%.

Moreover, the angular distribution of the $W$ and/or $Z$  bosons in the rest frame of the resonance with respect to the colliding direction can be used to diagnose the spin nature of the resonance.  The spin-$0$ resonance in our model would result in a uniform distribution, while models with spin-$1$ resonance should predict a parabolic distribution.

%%%%%%%%%%%%%%%%%%%%%%%%%%%%%%%%%%%%%%
\begin{figure}[t]
\begin{center}
  \includegraphics[width=.6\linewidth]{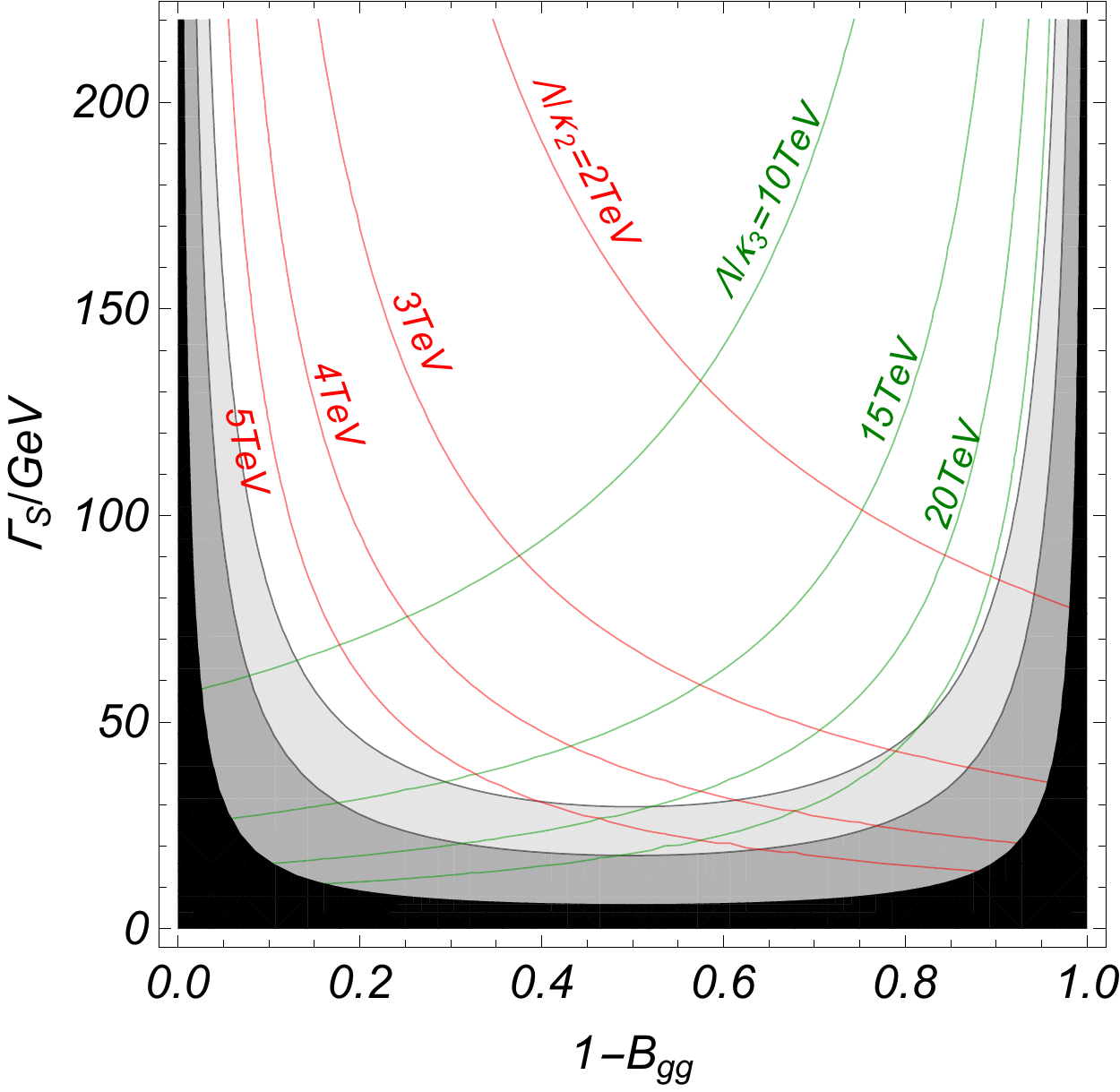}
 \end{center}
\caption{\sl \small
Contours of the coefficients $\Lambda/\k_{2,3}$ in Eq.~(\ref{eq:L3})
on the $(1-B_{gg})$-$\G_S$ plane at the 8-TeV LHC for $M_S = 2$\,TeV.
The dark, medium, and light shaded regions correspond to the cases where $\s_{WW} + \s_{ZZ} < 1$\,fb, $3$\,fb and $5$\,fb, respectively (see also Fig.\,\ref{fig:crosssection}). 
}
\label{fig:scales}
\end{figure}
%%%%%%%%%%%%%%%%%%%%%%%%%%%%%%%%%%%%%%

So far, we have not discussed the sizes of the coefficients $\kappa_i$ in Eq.\,(\ref{eq:SEFT}).
Since the decay widths are controlled by $M_S$ and the coefficients,  
we can estimate the required sizes of them as a function of $\G_S$ and $B_{gg}$.
In Fig.\,\ref{fig:scales}, we draw the contours of
\begin{eqnarray}
\label{eq:L3}
\frac{\Lambda}{\k_3} &=& \left(\frac{2 g_s^4 M_S^3}{\pi\G_S B_{gg}}\right)^{1/2}\ , \\
\label{eq:L2}
\frac{\Lambda}{\k_2} &\simeq& \left(\frac{3g^4 M_S^3}{4\pi\G_S(1- B_{gg})}\right)^{1/2}\ ,
\end{eqnarray}
on the plane of $B_{gg}$ and $\G_S$, thereby giving us some rough ideas about the corresponding dynamical scales (see Eqs.\,(\ref{eq:SGG}), (\ref{eq:SWW}),
(\ref{eq:SZZ}) and (\ref{eq:B})).
In Eq.~(\ref{eq:L2}), we have used the approximation that $g' \simeq 0$.
The figure shows that an appropriate cross section, $\s_{WW}+ \s_{ZZ} = {\cal O}(1$--$10)$\,fb,  is 
obtained for those parameters in the range of ${\cal O}(1$--$10)$\,TeV.
It is also noted that the parameter space with $B_{WW}+B_{ZZ} \sim 90$\% is disfavored
since a rather low suppression scale is required to explain the excesses (see also Eq.\,(\ref{eq:relation})). 
As these coefficients are intimately related to the dynamical scale of the hidden strong dynamics 
behind the scalar composite field $S$, this result suggests that the strong dynamics is also at around the TeV scale.

%%%%%%%%%%%%%%%%%%%%%%%%
\section{Hidden Strong Dynamics}
\label{sec:model}

In the above discussion, we have shown that the diboson excess events reported by the ATLAS
Collaboration can be explained by a spin-$0$ resonance with a mass of $2$\,TeV,
provided that the dimension-5 operators are suppressed by a mass scale of ${\cal O}(1$--$10$)\,TeV.
The proximity between the resonance mass and the suppression scale hints at the existence of strong dynamics with 
a dynamical scale $\Lambda_{\rm dyn}$ at around ${\cal O}(1)$\,TeV.
The narrow width of the resonance can be explained if it appears as the lightest composite state of the hidden dynamics.  In this section, we give an example whose lightest spin-$0$ composite field $S$
couples to the SM gauge bosons as in Eq.\,(\ref{eq:SEFT}).

%%%%%%%%%%%%%%%%%%%%%%%%%
\subsection{Composite scalar as the lightest state in hidden strong dynamics}

\begin{table}[t]
\caption{\sl \small
Charge assignments of the {bi-fundamental} scalars 
under the hidden $SU(5)$ and the SM gauge symmetries.
The SM gauge charges of the $Q$'s are assigned so that they form an anti-fundamental
representation of $SU(5)_{\rm GUT}$.
}
\begin{center}
\begin{tabular}{|c|c|c|c|c|c|}
\hline
&$SU(5)$&$SU(3)_c$&$SU(2)_{L}$ & $U(1)_Y$
\\
\hline
$Q_L$& ${\mathbf 3}$& ${\mathbf 1}$& ${\mathbf 2}$ & $1/2$
\\
$Q_D$& ${\mathbf 3}$& $\bar{\mathbf 3}$& ${\mathbf 1}$ & $-1/3$
\\
\hline
\end{tabular}
\end{center}
\label{tab:Q}
\end{table}

Let us start by considering a hidden $SU(N_c)$ gauge theory.
%If the hidden strongly interacting sector consists of a pure $SU(N_c)$ Yang-Mills theory,
%the lightest states below dynamical scale $\Lambda$ is expected to be 
%a scalar boson, the glueball $S$, with mass at around $\Lambda$.
%The glueball mass of $2$\,TeV is achieved by taking $\Lambda = {\cal O}(1)$\,TeV, although
%precise relation between the glueball mass and $\Lambda$ is hardly known.
The hidden dynamics is connected to the SM sector via a set of scalar fields $Q$'s that carry both the $SU(N_c)$ and the SM gauge charges.
The charge assignments of $Q$'s are given in Table.\,\ref{tab:Q}.
As an explicit example, we take $N_c = 5$ (see discussions at the end of this section), though most of the following discussions can be applied to different choices of the hidden gauge group. 
We assign the SM gauge charges to $Q$'s in such a way that they form an anti-fundamental
representation of the $SU(5)_{GUT}$ gauge group, the minimal $SU(5)$ grand unified theory (GUT).
In the following, $Q_{L,D}$ denote the bi-fundamental scalars.

Let us assume that the bi-fundamental scalars have masses, $m_{D,L}$, so that
\begin{eqnarray}
{\cal L} \supset -m_D^2 Q_D^\dagger Q_D - m_L^2  Q_L^\dagger Q_L\ .
\end{eqnarray}
When these masses are smaller than the dynamical scale, $m_{L,D} \lesssim \Lambda_{\rm dyn}$,
the lightest composite state is expected to be generally a mixture of 
composite mesons consisting of a pair of $Q$ and $Q^\dagger$ and a glueball. 
In our analysis, we assume that the lightest scalar state is dominated by the neutral meson states%
\footnote{The possibility of the glueball-dominated scenario is discussed in Appendix~\ref{sec:app}.}
\begin{eqnarray}
S \propto \cos\h_Q \times [Q_L^\dagger Q_L] +  \sin\h_Q\times  [Q_D^\dagger Q_D] \ ,
\end{eqnarray}
where $\h_Q$ parameterizes the relative contents of $Q_D^\dagger Q_D$ and $Q_L^\dagger Q_L$.
For example, the $Q_D^\dagger Q_D$ content is expected to be suppressed for $m_{D}\gg m_L$, although it is difficult to estimate $\theta_Q$ quantatively due to the non-perturbative nature of the interaction.%
\footnote{Here we naively assume that the lightest singlet scalar corresponds to the singlet under $SU(5)_{\rm GUT}$ in the limit of $m_D = m_L$.  If the lightest singlet scalar is dominated by the one in the adjoint representation of $SU(5)_{\rm GUT}$, on the other hand, $\tan\h_Q = -2/3$ even for $m_D = m_L$.}
In the following, we assume that the hidden strong dynamics does not cause spontaneous breaking of the SM gauge symmetries.

Using the Naive Dimensional Analysis (NDA)\,\cite{Cohen:1997rt,Luty:1997fk}, the scalar boson $S$ is matched to 
the composite fields by
\begin{eqnarray}
S \simeq 
\frac{4\pi}{\k\L_{\rm dyn}}\cos\h_Q \times [Q_L^\dagger Q_L] 
+ \frac{4\pi}{\k\L_{\rm dyn}}\sin\h_Q\times  [Q_D^\dagger Q_D] \ ,
\end{eqnarray}
where $\k$ is an ${\cal O}(1)$ coefficient within the uncertainty of the NDA.
As a result, we obtain the effective interactions of $S$ to the SM gauge bosons as
\begin{eqnarray}
\label{eq:SEFT3}
{\cal L}_{\rm eff} &=&
\frac{\k}{4\pi\L_{\rm dyn}}\sin\h_Q
\,
S G^a_{\mu\nu}G^{a\,\mu\nu}
 + 
 \frac{\k }{4\pi\L_{\rm dyn}}\cos\h_Q\,
 SW^i_{\mu\nu}W^{i\,\mu\nu}
\nonumber \\
&& +
 \frac{2\k}{4\pi\L_{\rm dyn}}
\left(
\frac{\sin\h_Q}{3} 
+
 \frac{\cos\h_Q}{2}
\right)
S
B_{\mu\nu}B^{\mu\nu}\ .
\label{eq:GGGG2}
\end{eqnarray}
%where $c$ is a numerical factor which is common for all the Standard Model gauge groups.
By comparing with Eq.\,(\ref{eq:SEFT}), we can then identify the coefficients used in 
the previous section:
\begin{eqnarray}
\label{eq:relation}
\frac{\k_3}{\L} = \frac{\k\sin\h_Q}{4\pi \L_{\rm dyn}}\ , \,\,\,\,
\frac{\k_2}{\L} = \frac{\k\cos\h_Q}{4\pi \L_{\rm dyn}}\ , \,\,\,\,
\frac{\k_1}{\L} =  \frac{\k}{4\pi \L_{\rm dyn}}
 \frac{6}{5}
\left(
\frac{\sin\h_Q}{3} 
+
 \frac{\cos\h_Q}{2}
\right)\ . 
\end{eqnarray}

As discussed in the previous section, the scales $\Lambda/\k_{1,2,3}$ are required to be of ${\cal O}(1$--$10)$\,TeV to account for the diboson excess (see Fig.\,\ref{fig:scales}).  On the other hand, the mass of $S$, $M_S\simeq 2$\,TeV, is expected to be of ${\cal O}(\Lambda_{\rm dyn})$.
These conditions are simultaneously satisfied for $\k \sim O(1)$, consistent with the NDA.

%%%%%%%%%%%%%%%%%%%%%%%%%%%%%%%%%%%%%%
\begin{figure}[t]
\begin{center}
  \includegraphics[width=.6\linewidth]{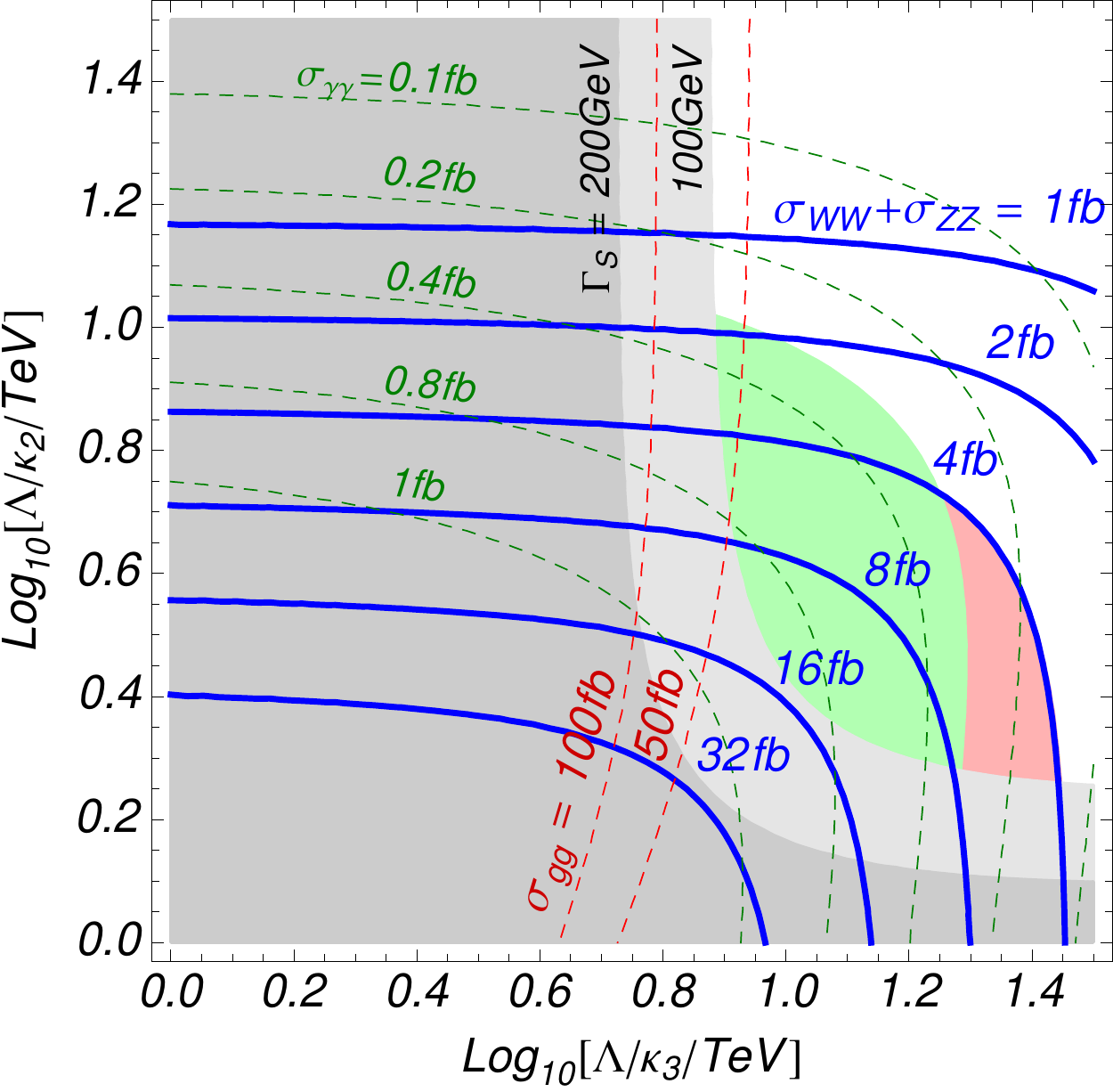}
 \end{center}
\caption{\sl \small Contours of various diboson processes at the 8-TeV LHC on the $\L/\k_2$-$\L/\k_3$ plane for $M_S = 2$\,TeV.  Blue curves give different production cross sections of $S$ decaying into $WW$ and $ZZ$ final states.
Red dashed curves are contours of the cross section of the digluon mode that contributes to dijets.
Green dashed curves are contours of the cross section of the diphoton mode.  Boundaries of the gray regions have fixed total widths ($100$ and $200$ GeV) for the resonance.  
The light green region is excluded by the constraints on the diphoton channel, $\s_{\g\g}\lesssim 0.3$\,fb\,\cite{Aad:2015mna}.}
\label{fig:meson}
\end{figure}
%%%%%%%%%%%%%%%%%%%%%%%%%%%%%%%%%%%%%%

In Fig.\,\ref{fig:meson}, we show various contours on the plane of $\L/\k_2$ and $\L/\k_3$, the two of which 
are related to $\k/(4\pi \L_{\rm dyn})$ and $\theta_Q$ via Eq.\,(\ref{eq:relation}).%
\footnote{The effective field theory is controlled by two parameters, $\k/(4\pi\L_{\rm dyn})$ and $\theta_Q$, in this dynamical model.
In particular, the branching ratio of each $S$ decay mode is solely determined by $\theta_Q$.}
The figure reconfirms that the cross section $\sigma_{WW} + \sigma_{ZZ} = {\cal O}(1$--$10)$\,fb
is achieved for $\L/\k_{2} = {\cal O}(1)$\,TeV and $\L/\k_{3} = {\cal O}(1$--$10)$\,TeV,
while keeping the total width of $S$ sufficiently narrow, as indicated by the green region.%
\footnote{The total cross section $\s_{WW} + \s_{ZZ}$ is slightly smaller than the one shown in Fig.\,\ref{fig:crosssection}, where we have neglected the $\g\g$ and $Z\g$ modes.} 
As the figure shows, it is preferred to have a smaller value for $\k_3/\k_2 \simeq \tan\h_Q$.
This can be readily achieved when the mass of $Q_D$ is larger than that of $Q_L$.

In the figure, we also show contours of the cross section of the diphoton channel, $\s_{\g\g}$, which is about $5$--$10$\% of $\s_{WW} + \s_{ZZ}$.  
The light green region is excluded by the constraints on the diphoton channel, $\s_{\g\g}\lesssim 0.3$\,fb\,\cite{Aad:2015mna},
which is one of the most constraining channels at LHC Run-I.
By remembering that the production cross section of $S$ is enhanced by a factor of ten at LHC Run-II (see Eq.\,(\ref{eq:13TeV})), 
it is possible to test this model by searching for the diphoton signals.

So far, we have not included couplings between the bi-fundamental scalars and the Higgs bosons, such as,
\begin{eqnarray}
\label{eq:QQHH}
{\cal L} =  \lambda \left( H^\dag \G^A H \right) \left( Q_{L,D}^\dag \G^A Q_{L,D} \right) ~,
\end{eqnarray}
where $\l$ represents a coupling constant and 
$\G^A = 1$ or the Pauli matrix, $\G^A = \sigma^i$, for $Q_L$ and $\G^A = 1$ for $Q_{D}$.%
If we allow such interactions, the resonance also decay into a pair of Higgs bosons,
which alter the total decay width of $\G_S$ as well as the branching ratios, in particular the ratio of 
the diphoton mode.
So far, the resonance decay into pair of the higgs is not severely constrained\,\cite{Aad:2015uka}. 
In this paper, we simply assume that the direct couplings between $Q$'s and the Higgs bosons are 
somewhat suppressed.

We also comment on the constraints from electroweak precision measurements.
The most dangerous effect is from the interaction term in Eq.\,(\ref{eq:QQHH}) with $\G^A = \s^i$ 
for $Q_L$.
Below the dynamical scale of the hidden strong dynamics, we obtain the effective interaction
\begin{eqnarray}
{\cal L} \simeq \frac{\lambda}{4\pi} \Lambda_{\rm dyn} H^\dag \sigma^i H T^i ~,
\end{eqnarray}
where $T^i$ is the composite triplet scalar.
After the Higgs field obtains a vacuum expectation value $v_{\rm EW}$, 
the triplet is also induced to have a vacuum expectation value,
\begin{eqnarray}
\vev{T^3} \simeq \frac{\lambda v_{\rm EW}^2  \Lambda_{\rm dyn}}{4\pi M_T^2} = 0.6~{\rm GeV} \times \lambda\, \frac{\Lambda_{\rm dyn}}{1~{\rm TeV}} \left(\frac{M_T}{2~{\rm TeV}} \right)^{-2} ~,
\end{eqnarray}
where $M_T$ is the mass of the triplet.
As long as $\lambda \lsim O(1)$, the constraint from the $T$ parameter can be evaded.
Contributions to the $S$, $T$, $U$ parameters by quantum corrections~\cite{Peskin:1991sw} are also suppressed by the dynamical scale and hence small.

%%%%%%%%%%%%%%%%%%%%%%%%%
\subsection{Charged composite states, dark matter candidate}

In the previous sections, we have concentrated exclusively on the production of the lightest neutral spin-$0$ boson. 
In addition to the neutral scalar $S$, the dynamical model also predicts
scalar particles charged under the SM gauge symmetries: an $SU(3)_c$ octet, an $SU(2)_L$ triplet,
and a bi-fundamental representation of $SU(3)_c\times SU(2)_L$ with 
a $U(1)_Y$ charge of $5/6$.

The octet scalar is pair produced via QCD processes and singly produced via dimension-5 operators coupling to the gluons.
So far, the production cross section of the octet scalar is constrained to be smaller than about
$100$\,fb for a mass around 2 TeV~\cite{Aad:2014aqa,Khachatryan:2015sja}.
In both production processes, this constraint is evaded.

On the other hand, the triplet scalar is produced via the Drell-Yan process
and immediately decays into SM electroweak gauge bosons and Higgs bosons through the interaction in Eq.\,(\ref{eq:QQHH}).%
\footnote{The mass of the triplet is expected to be larger than that of $S$, 
since $S$ is a mixture of $Q_L^\dagger Q_L$ and $Q_D^\dagger Q_D$.}
Unlike the neutral scalar $S$, the triplet scalar does not couple to the 
gluons via any dimension-5 operator.
Up to date, there is no stringent constraint on the triplet scalar with a mass of ${\cal O}(1)$\,TeV.

The scalar of bi-fundamental representation of $SU(3)_c\times SU(2)_L$  requires a special care,
as it cannot decay into a pair of SM gauge bosons.
In order for it to decay promptly, we introduce a pair of fermions $(\psi_Q, \bar{\psi}_Q)$ which are the fundamental and the anti-fundamental representations of the hidden $SU(5)$ gauge symmetry.
With these fermions, the bi-fundamental scalars $Q_{D,L}$ in the dynamical model couple 
to the SM quarks and leptons, $\bar{d}_R$  and $\ell_L$, via
\begin{eqnarray}
{\cal L} \supset y\, Q_D^\dagger\, \psi_Q \, \bar{d}_R + 
y\, Q_L^\dagger\, \psi_Q \, \ell_L + 
M \psi\bar{\psi} \ ,
\end{eqnarray}
where $y$ denotes some coupling constant and $M$ denotes the mass of the fermion $\psi_Q$.%
\footnote{By taking $M$ much larger than a TeV, these additional fermions cannot be produced at the LHC.}
Through these interactions, the $Q_D^\dagger Q_L$ bound states immediately decay 
into a pair of $\bar d_R$ and $\ell_L$.
With a sufficiently short lifetime, there is no stringent constraint on the
bi-fundamental representation of $SU(3)_c\times SU(2)_L$ with a mass of ${\cal O}(1)$\,TeV.

Before closing this section, let us comment on the baryonic states of the hidden $SU(5)$ gauge interaction.
The lightest baryonic scalar is given by,
\begin{eqnarray}
B \propto  QQQQQ\ ,
\end{eqnarray}
which is neutral under the SM gauge groups.
This neutrality of $B$ is the reason why we have chosen $N_c = 5$ for the hidden strong gauge interaction.
It should be noted that the lightest baryonic state is stable
due to an approximate $U(1)$ symmetry.%
\footnote{Here we assume that the $U(1)$ symmetry is not spontaneously broken by the strong gauge dynamics.}
Therefore, the baryonic scalar serves as a good candidate for dark matter.

At the early universe, the baryonic scalars annihilate into a pair of light scalar composite fields. 
The thermal relic abundance is expected to be much lower than the observed dark matter density if the annihilation cross section saturates the unitarity limit~\cite{Griest:1989wd}. 
However, the mass of the lightest baryonic scalar is higher than the dynamical scale.
Thus, the effective coupling between the light scalar composites and  baryon dark matter  
can be suppressed by form factors, which leads to a somewhat suppressed annihilation cross section.
If this is the case, the observed dark matter density may be explained 
by the thermal relic density of the baryonic dark matter in the model.

The strong dynamics also predicts heavier composite modes.
The heavier composite states are expected to decay into the light scalar composite 
or the lightest baryon state by emitting $S$, $\bar{d}_R$ and $\ell_L$, so that there is no stringent constraint on them.

\section{Conclusions and Discussions}

In this paper, we have proposed a new explanation for the excess events observed in the search
for a high-mass resonance decaying into dibosons by the ATLAS experiment.
The resonance is identified as a {\it composite} spin-$0$ particle coupling 
to the Standard Model gauge bosons via dimension-5 operators.
We find that the reported excess can be explained if the dimension-5 operators 
are suppressed by a mass scale of ${\cal O}(1$--$10$)\,TeV.
As a notable feature of our model, the resonance decays into a pair of photons, which is absent in proposals of interpreting the resonance as a spin-$1$ particle.

We have also constructed a model of hidden strong dynamics which realizes the spin-$0$ particle as its lightest composite state, with appropriate couplings to the Standard Model gauge bosons.
In this scenario, the composite spin-$0$ particle consists of bi-fundamental scalars of the hidden and the Standard Model gauge symmetries.
The mass of the resonance as well as the suppression scale of ${\cal O}(1$--$10)$\,TeV
are achieved when the hidden strong dynamics exhibits confinement at a dynamical scale of ${\cal O}(1)$\,TeV.
Along with the neutral scalar boson, the dynamical model predicts many charged particles
whose masses are also in the TeV regime.
Therefore, we expect in this model that the LHC Run-II experiment will discover a zoo of particles around that scale.

A natural question about the diboson resonance at the TeV scale is ``who ordered that?''
One possible answer is the dark matter.  In our model, for example, there is a dark matter candidate in the hidden sector with a mass in the TeV regime.
In conjunction with the anthropic arguments, the dynamical scale at the TeV regime may be justifiable.
If the scale of the resonance is related to the origin of the electroweak scale, the TeV scale of the resonance may again be justifiable by anthropic arguments.

If, on the other hand, the dibsoson resonance is not directly related to either the dark matter or the electroweak scale, the resonance at the TeV scale provides a strong counterexample to the anthropic arguments.
In such a case, the TeV scale of the resonance needs to be explained for its own sake by, for example, supersymmetry.
Interestingly, the dynamical model in section\,\ref{sec:model} has almost an identical structure to the supersymmetric model in Ref.~\cite{Evans:2012uf}, where the dynamical sector
was introduced to achieve the observed Higgs boson mass in the MSSM
with soft supersymmetry breaking masses in the TeV regime.
We will discuss the diboson resonance in the supersymmetric model in a separate work.

%%%%%%%%%%%%%%%%%%%%%%%%%%%%%%%%%%%%%%%%%%%%%%%%
\headline{Acknowledgements}\\
We thank Satoshi Shirai for useful discussion.
This work is supported in part by the Ministry of Science and Technology of Taiwan under Grant No.~MOST-100-2628-M-008-003-MY4 (C.-W.~C), Grants-in-Aid for Scientific Research from the Ministry of Education, Culture, Sports, Science, and Technology (MEXT), Japan, No. 24740151 and No. 25105011 (M.~I.) as well as No. 26104009 (T.~T.~Y.); 
Grant-in-Aid No. 26287039 (M.~I. and T.~T.~Y.) from the Japan Society for the Promotion of Science (JSPS); and by the World Premier International Research Center Initiative (WPI), MEXT, Japan (H.~F., M.~I., and T.~T.~Y.).
K.H. is supported in part by a JSPS Research Fellowship for Young Scientists.
%
%---------------SECTION------------------%

\appendix
\section{The composite scalar S as a glueball in hidden strong dynamics
\label{sec:app}}
In our discussion in section~\ref{sec:model}, 
we identify the lightest scalar boson in the hidden sector with the lightest 
meson consisting of $Q$ and $Q^\dagger$.
If we take the mass parameter of $Q$'s larger than the dynamical scale, on the other hand,
the lightest state is expected to be dominated by a scalar glueball in the hidden sector.
In this appendix, we discuss whether the glueball can be a good candidate of the diboson resonance.

When the masses of the bi-fundamental scalars are heavier than $\Lambda_{\rm dyn}$,
we may integrate out the bi-fundamental scalars.
At the leading order, the effective interactions between the $SU(5)$ gauge bosons and Standard Model 
gauge bosons are given by~\cite{Novikov:1977dq},
\begin{eqnarray}
\label{eq:SEFT2}
{\cal L}_{\rm eff} &\simeq&
\frac{1}{144\cdot16\pi^2}
 \frac{1}{m_{D}^4} 
 H_{A\mu\nu} H^{A\,\mu\nu} 
 G^a_{\mu\nu}G^{a\,\mu\nu}
 + 
 \frac{1}{144\cdot16\pi^2}
 \frac{1}{m_{L}^4} 
  H^A_{\mu\nu} H^{A\,\mu\nu}  W^i_{\mu\nu}W^{i\,\mu\nu}\nonumber \\
&& + 
 \frac{2}{144\cdot16\pi^2}
\left(
\frac{1}{3} \frac{1}{m_{D}^4} 
+
\frac{1}{2} \frac{1}{m_{L}^4} 
\right)
H_{A\mu\nu} H^{A\,\mu\nu}  B_{\mu\nu}B^{\mu\nu}\ ,
\label{eq:GGGG}
\end{eqnarray}
where $ H^A_{\mu\nu}$ ($A =1-24$) denote the field strengths of $SU(5)$ gauge bosons, which are again normalized to have
\begin{eqnarray}
{\cal L } \supset - \frac{1}{4g_H^2}  H^A_{\mu\nu}H^{A\,\mu\nu} \ .
\end{eqnarray}

Below the mass scale of $Q$'s, the hidden sector ends up with a strongly interacting pure Yang-Mills theory. 
At around the dynamical scale $\Lambda_{\rm dyn}$, the gauge coupling constant in the hidden sector becomes strong, {\it i.e.}, $g_H \sim 4\pi$, and confinement is expected to occur.
In this case, the scalar glueball $S$ becomes the lightest state.
To match the scalar glueball $S$ to $HH$ in Eq.\,(\ref{eq:GGGG}), we again use the NDA:
\begin{eqnarray}
\label{eq:NDA0}
H_{A\mu\nu} H^{A\,\mu\nu} = 4\pi \k  \,\Lambda_{\rm dyn}^3 S\ .
\end{eqnarray}
By substituting Eq.\,(\ref{eq:NDA0}) into Eq.\,(\ref{eq:GGGG}), we obtain the effective interactions between the glueball and SM gauge bosons,
\begin{eqnarray}
\label{eq:SEFT2}
{\cal L}_{\rm eff} &=&
\frac{\k }{72\pi}
 \frac{\L_{\rm dyn}^3}{m_{D}^4} 
S
 G^a_{\mu\nu}G^{a\,\mu\nu}
 + 
 \frac{\k }{72\pi}
 \frac{\L_{\rm dyn}^3}{m_{L}^4} 
  S
  W^i_{\mu\nu}W^{i\,\mu\nu}
+ 
 \frac{\k}{72\pi}
\left(
\frac{2}{3} \frac{\L_{\rm dyn}^2}{m_{D}^4} 
+
 \frac{\L_{\rm dyn}^3}{m_{L}^4} 
\right)
S
B_{\mu\nu}B^{\mu\nu}\ .
\label{eq:GGGG2}
\end{eqnarray}
By comparing this with Eq.\,(\ref{eq:SEFT}), we find
\begin{eqnarray}
\label{eq:relation2}
\frac{\k_3}{\L} = \frac{\k}{72\pi}\frac{\L_{\rm dyn}^3}{m_D^4}\ ,\,\,\,\,
\frac{\k_2}{\L} = \frac{\k}{72\pi}\frac{\L_{\rm dyn}^3}{m_L^4}\ ,\,\,\,\,
\frac{\k_1}{\L} =  \frac{\k}{72\pi}
 \frac{6}{5}
\left(
\frac{1}{3} \frac{\L_{\rm dyn}^3}{m_{D}^4} 
+
\frac{1}{2}
 \frac{\L_{\rm dyn}^3}{m_{L}^4} 
\right)\ .
\end{eqnarray}
%where $S$ is the scalar glueball. 

As discussed in section\,\ref{sec:SEFT}, the scales $\Lambda/\k_{1,2,3}$ 
are required to be ${\cal O}(1-10)$\,TeV to account for the reported diboson excess.
On the other hand, the mass of the glueball, $M_S\simeq 2$\,TeV, is expected to be of ${\cal O}(\Lambda_{\rm dyn})$.
To satisfy these conditions, the relations in Eq.\,(\ref{eq:relation}) requires $\k = {\cal O}(100)$
even for a rather small mass of $Q$'s, {\it i.e.},~$m_{L} \sim m_{D} \sim \Lambda_{\rm dyn}$.
Such a large $\k$ seems contradicting with the NDA expectation. Therefore, it is unlikely that
the excess can be explained by the glueball state.

\end{document}